\documentclass[runningheads]{llncs}
\usepackage{graphicx}
\usepackage{amsmath}
\usepackage{tabularx}
\usepackage{multirow}

\usepackage[table]{xcolor}

\definecolor{baselinecolor}{gray}{.9}

\definecolor{Gray}{gray}{0.9}

\usepackage{algorithm}
\usepackage{algpseudocode}
\usepackage{amsmath}

\begin{document}
\title{LatentEdit: Adaptive Latent Control for Consistent Semantic Editing}

\author{
Siyi Liu\inst{1}\thanks{Equal contributions}\orcidID{0009-0001-4263-535X} \and
Weiming Chen\inst{1\star}\orcidID{0000-0002-0586-1278} \and
Yushun Tang\inst{1}\orcidID{0000-0002-8350-7637} \and
Zhihai He\inst{1,2}\thanks{Corresponding author}\orcidID{0000-0002-2647-8286}
}
\authorrunning{S. Liu, W. Chen et al}
\institute{Southern University of Science and Technology, China \and
Pengcheng Laboratory, China \\
\email{\{12332140,chenwm2023,tangys2022\}.mail.sustech.edu.cn} \\
\email{hezh@sustech.edu.cn}}
%
% \institute{}
\maketitle              % typeset the header of the contribution
\begin{abstract}
Diffusion-based Image Editing has achieved significant success in recent years. However, it remains challenging to achieve high-quality image editing while maintaining the background similarity without sacrificing speed or memory efficiency. In this work, we introduce LatentEdit, an adaptive latent fusion framework that dynamically combines the current latent code with a reference latent code inverted from the source image. By selectively preserving source features in high-similarity, semantically important regions while generating target content in other regions guided by the target prompt, LatentEdit enables fine-grained, controllable editing. Critically, the method requires no internal model modifications or complex attention mechanisms, offering a lightweight, plug-and-play solution compatible with both UNet-based and DiT-based architectures. Extensive experiments on the PIE-Bench dataset demonstrate that our proposed LatentEdit achieves an optimal balance between fidelity and editability, outperforming the state-of-the-art method even in 8-15 steps. Additionally, its inversion-free variant further halves the number of neural function evaluations and eliminates the need for storing any intermediate variables, substantially enhancing real-time deployment efficiency.

\end{abstract}

\begin{figure}
\includegraphics[width=\textwidth]{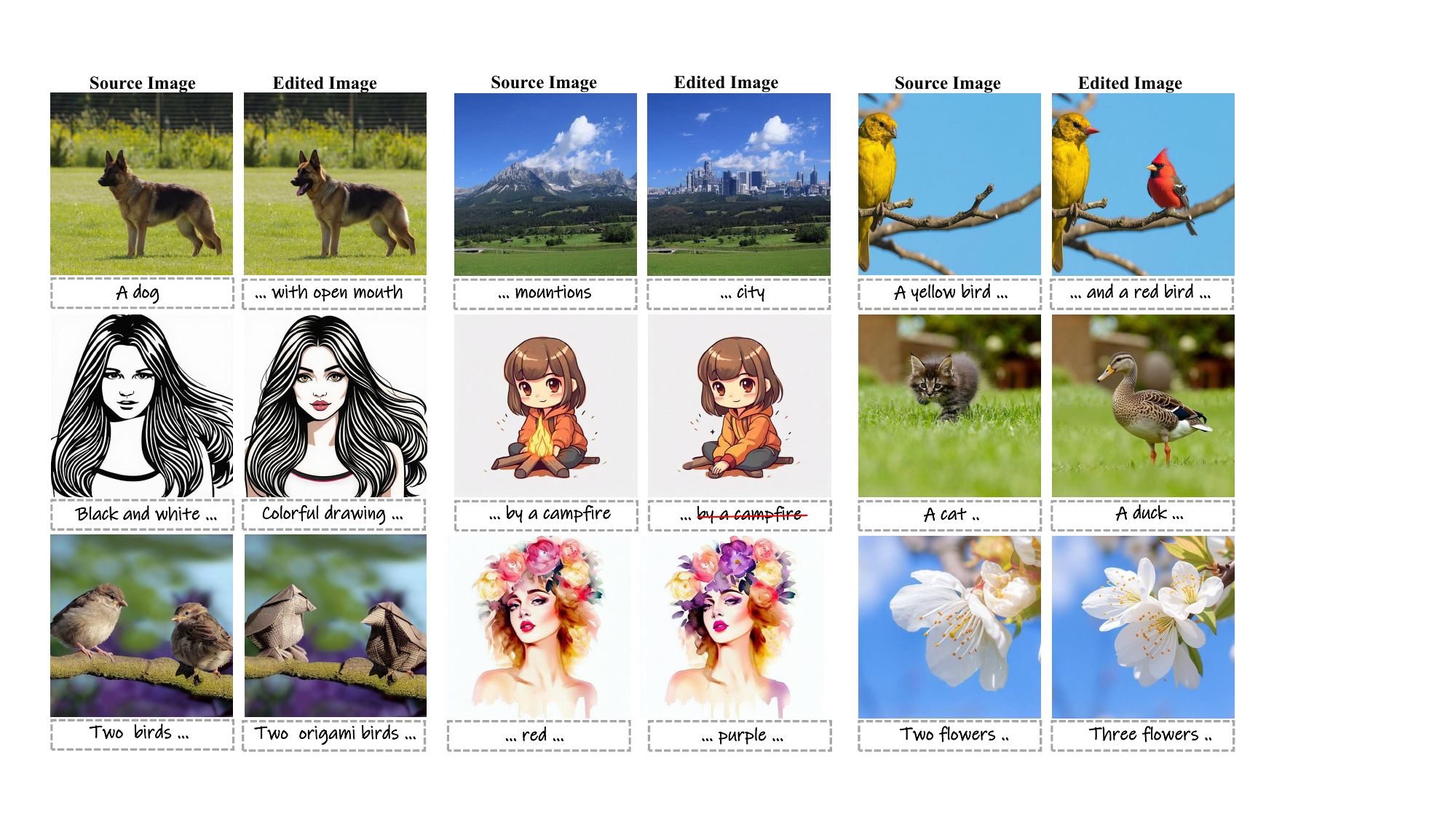}
\caption{LatentEdit for real image editing. Our method delivers strong performance across diverse editing tasks, achieving precise text-image alignment while suppressing unintended changes.} 
\label{introduction}
\end{figure}

\section{Introduction}
Recent advances in diffusion-based generative models~\cite{Ho2020DDPM,Ramesh2022DALL-E2,Saharia2022Imagen,Rombach2022LDM,Balaji2023eDiff-I,Betker2023DALL-E3,Podell2024SDXL,Sauer2024SDXLTurbo,Chen2024PixArt-alpha,Esser2024RectifiedFlowTransformers} have significantly transformed the field of text-to-image generation. These models synthesize high-quality images by progressively denoising Gaussian noise, guided by textual prompts provided by users. Among various models, Stable Diffusion (SD)~\cite{Crowson2022StableDiffusion}, which adopts UNet architecture and DDIM~\cite{Song2021DDIM} sampling strategy, as well as FLUX~\cite{BlackForestLab2024FLUX}, which employs Multimodal Diffusion Transformer (MM-DiT)~\cite{Peebles2023DiT} with Rectified Flow~\cite{Liu2023FlowMatching,Lipman2023FlowMatching,Esser2024RectifiedFlowTransformers} sampling method, are two of the most widely used models due to their powerful generative capabilities. 

Researchers are eager to leverage the powerful generative capability of these models to manipulate real-world images. Therefore, the key challenge is \textit{how to manipulate a real-world image while preserving its style or semantic content}. Previous works~\cite{Hertz2023P2P,Cao2023MasaCtrl,Tumanyan2023PnP,Wang2024RF-Solver,Deng2024FireFlow,Zhu2025KV-Edit} leverage model internal features from the inversion process to improve the consistency of the edited image. However, directly fusing high-dimensional internal features may introduce conflicts within the model, potentially leading to performance degradation. Moreover, storing these features incurs substantial memory overhead. This motivates us to develop a more effective and efficient editing approach.

In this work, we propose \textbf{LatentEdit}, a novel and efficient approach that performs adaptive fusion directly in the latent space. Instead of manipulating complex attention features or modifying internal layers of the model, we guide the denoising process by measuring the spatial similarity between the current latent and a reference latent chain extracted from the source image. This allows for fine-grained control that selectively retains content in semantically important regions while allowing the prompt to drive change in others. Our approach is lightweight, compatible with both inversion-based and inversion-free pipelines, and seamlessly applicable to both UNet-based and DiT-based architectures. We demonstrate through extensive experiments that LatentEdit achieves state-of-the-art performance with superior consistency and efficiency across a range of image editing tasks.

\section{Related Work and Unique Contributions}

In this section, we first review diffusion inversion methods that connect real-world images with diffusion models. Then, we review the existing text-guided image editing approaches related to this work. Finally, we summarize the unique contributions of this work.

\subsection{Diffusion Inversion Methods}

Inversion bridges real-world images and text-to-image diffusion models by inverting a given image back to a specific Gaussian noise sample, such that the diffusion model can reconstruct the image through denoising. Therefore, inversion serves as a basic building block for real-world image manipulation. Existing inversion methods can be categorized into two major types based on the sampling method: DDIM-based and RF-based approaches.

Among existing text-to-image diffusion models, Stable Diffusion (SD)~\cite{Crowson2022StableDiffusion} is the most commonly used open-source model, which relies on the DDIM sampling method~\cite{Song2021DDIM}. Existing DDIM-based inversion methods can be categorized into 4 types: deterministic, numerical, tuning-based, and other methods. Deterministic methods~\cite{Song2021DDIM,Dhariwal2021DDIM,Miyake2023NPI} achieve inversion based on the reversible assumption of ordinary differential equations (ODEs). Numerical methods~\cite{Wallace2023EDICT,Pan2023AIDI,Samuel2025GNRI,Zhang2024BDIA,Garibi2024ReNoise,Wang2024BELM} employ numerical optimization techniques to provide more accurate approximations. Tuning-based methods~\cite{Mokady2023NTI,Dong2023PTI,Hong2024ExactDPM} achieve exact inversion by training some variables. Other methods~\cite{Zhang2023RIVAL,Duan2024TIC} reuse the features from the inversion process to align the sampling and inversion processes.

Due to the theoretical differences between DDIM and RF, the above DDIM-based methods cannot be directly applied to RF-based models (\textit{e.g.}, FLUX~\cite{BlackForestLab2024FLUX}). RF-Prior~\cite{Yang2025iRFDS} performs score distillation to invert a given image using RF models. RF inversion~\cite{Rout2025RF-Inversion} improves the inversion quality by employing dynamic optimal control derived from linear quadratic regulators. RF-Solver~\cite{Wang2024RF-Solver} uses the Taylor expansion to reduce inversion errors in the ODE process of RF models. FireFlow~\cite{Deng2024FireFlow} reuses intermediate velocity approximations to achieve the second-order accuracy while maintaining the computational cost of a first-order method.

To highlight the effectiveness of the proposed editing method, in this paper, we adopt the simplest inversion methods: DDIM inversion for SD and vanilla RF for FLUX. Notably, our inversion-free variant also achieves comparable performance to state-of-the-art methods with only the sampling branch required.

\subsection{Text-Guided Semantic Editing}

The goal of image editing is to modify the visual content in a controllable manner while ensuring consistency with the original image. Text-guided semantic editing modifies an image solely by changing the textual prompt and has attracted the most attention due to its flexibility~\cite{Huang2025EditSurvey}. 

Text-guided semantic editing has been widely studied for UNet-based models (\textit{e.g.}, SD). Prompt-to-Prompt (P2P)~\cite{Hertz2023P2P} injects attention maps from the inversion process to the sampling process to preserve the spatial layout and geometric structure of the original image. MasaCtrl~\cite{Cao2023MasaCtrl} introduces a mask-guided mutual self-attention mechanism, which replaces the key and value attention features in self-attention layers to enhance the consistency of the edited image. Plug-and-Play (PnP)~\cite{Tumanyan2023PnP} enables fine-grained control over generated structures by manipulating spatial and self-attention features, directly injecting features from a guidance image.

Due to the significant architecture difference between UNet-based (\textit{e.g.}, SD) and DiT-based (\textit{e.g.}, FLUX) models, the above methods fail to be applied to DiT-based models directly. RF-Solver~\cite{Wang2024RF-Solver} and FireFlow~\cite{Deng2024FireFlow} replace the value attention features in single-stream DiT blocks to balance the trade-off between fidelity and editability. 

Unlike previous methods that manipulate high-dimensional internal features, our approach performs adaptive fusion directly in the latent space, achieving superior performance without introducing burdensome computational overhead.

\subsection{Unique Contributions}

Compared to existing approaches, our unique contributions include: (1) We propose an efficient zero-shot text-guided image editing approach that ensures high consistency by adaptively fusing the original and edited latent representations. (2) Since our method does not manipulate internal model features, it serves as a plug-and-play solution that is compatible with both DDIM-based models (\textit{e.g.}, SD) and RF-based models (\textit{e.g.}, FLUX). (3) Our method is one of the fastest text-guided image editing approaches due to its tuning-free nature and avoidance of operating complex internal model features. Notably, our inversion-free variant reduces the number of Neural Function Evaluations (NFEs) by half while achieving consistency comparable to State-of-The-Art (SoTA) methods. (4) Extensive experimental results on the PIE-Bench dataset demonstrate that the proposed method achieves state-of-the-art performance on both fidelity and editability.

\section{Proposed Method}

In this section, we first review the background knowledge and present an overview of the proposed method. Then, we provide a detailed description of the proposed adaptive latent fusion method. Finally, we introduce the inversion-free variant of our method.

\subsection{Preliminaries and Method Overview}

\begin{figure}[t]
\includegraphics[width=\textwidth]{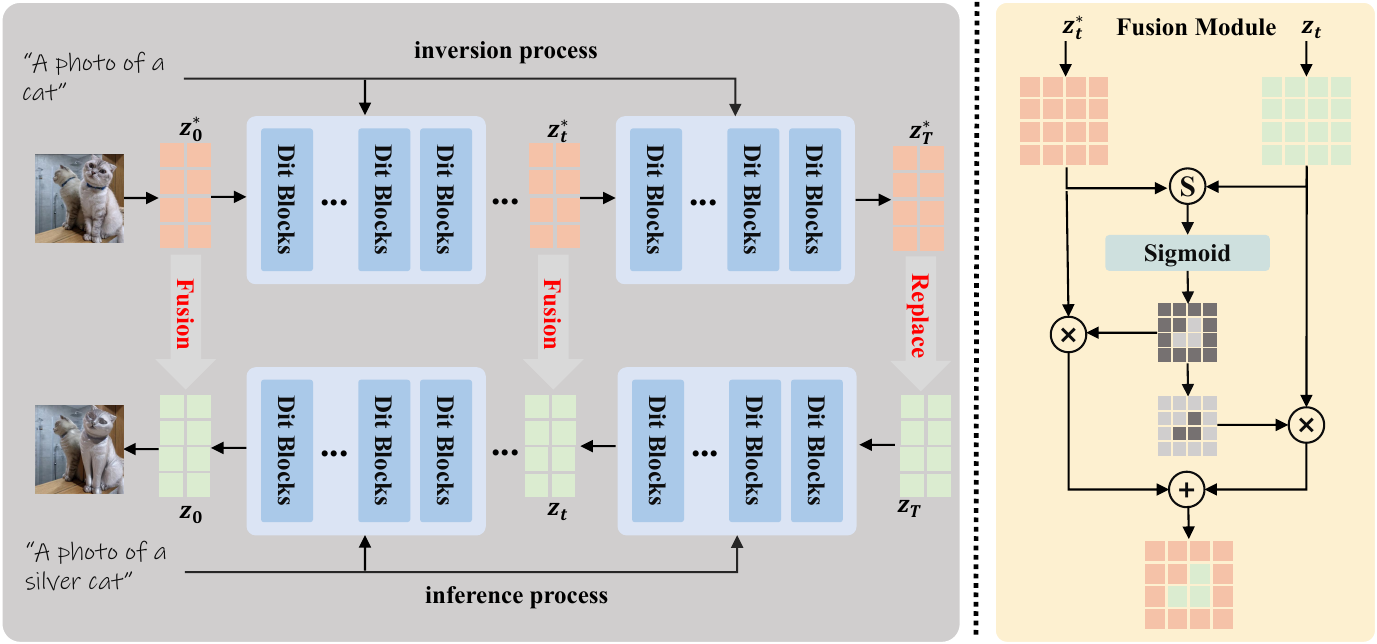}
\caption{Overview of the proposed LatentEdit. Given an input image $I^*$, we obtain a reference latent chain. During denoising, we dynamically compare the current latent $z_t$ with $z_t^*$; regions of high similarity retain source features, while dissimilar regions are guided by the target prompt. This enables content-consistent image synthesis aligned with the target prompt.} 
\label{fig1}
\end{figure}

\subsubsection{Preliminaries}
In text-to-image diffusion models, the forward pass adds noise to the image latent representation $z_0$. For DDIM-based models, the forward pass is defined as follows:
\begin{equation}
z_t = \sqrt{\bar{\alpha}_t} z_0 + \sqrt{1 - \bar{\alpha}_t} \epsilon,
\label{eq: DDIM add noise}
\end{equation}
where $\bar{\alpha}_t$ is the parameter at the t-$th$ timestep predefined by the DDIM sampler and $\epsilon \sim \mathcal{N} (0, \textbf{I})$ denotes the noise randomly sampled from the standard Gaussian distribution. The forward pass of RF models follows a linear path defined as:
\begin{equation}
z_t = t \epsilon + (1-t) z_0.
\label{eq: RF add noise}
\end{equation}
The text-to-image diffusion models gradually generate the image following a backward pass. The backward pass of DDIM-based models is defined as:
\begin{equation}
z_{t-1} = \sqrt{\frac{\bar{\alpha}_{t-1}}{\bar{\alpha}_t}} z_{t} + \left ( \sqrt{1 - \bar{\alpha}_{t-1}} - \sqrt{\frac{(1-\bar{\alpha}_{t}) \bar{\alpha}_{t-1}}{\bar{\alpha}_{t}}} \right ) \mathbf{F}_{\theta} \left ( z_t, t \right ).
\label{eq: DDIM denoise}
\end{equation}
As stated in the literature~\cite{Lu2022DPM-Solver}, sampling from diffusion models can alternatively be as solving the corresponding ODEs. Therefore, the sampling process can be reversed under the assumption that the ODE process is reversible in the limit of small steps:
\begin{equation}
z_t = \sqrt{\frac{\bar{\alpha}_t}{\bar{\alpha}_{t-1}}} z_{t-1} + \left ( \sqrt{1 - \bar{\alpha}_t} - \sqrt{\frac{(1-\bar{\alpha}_{t-1})\bar{\alpha}_t}{\bar{\alpha}_{t-1}}} \right ) \mathbf{F}_{\theta} (z_{t-1}, t-1).
\label{eq: DDIM inversion}
\end{equation}
As for the RF models, the transition between noise and data distributions is modeled by an ODE over a continuous time interval $t \in [0, 1]$: $dz_t = \mathbf{V} (z_t, t) dt$. In practice, the ODE is discretized and solved using the Euler method:
\begin{equation}
z_{t_{i-1}} = z_{t_{i}} + \left ( t_{i-1} - t_i \right ) \mathbf{V}_{\theta} \left ( z_{t_{i}}, t_{i} \right ).
\label{eq: RF denoise}
\end{equation}
Therefore, the vanilla inversion for RF models can be denoted as:
\begin{equation}
z_{t_{i}} = z_{t_{i-1}} + \left ( t_{i} - t_{i-1} \right ) \mathbf{V}_{\theta} \left ( z_{t_{i-1}}, t_{i-1} \right ).
\label{eq: vanilla RF inversion}
\end{equation}

\subsubsection{Method Overview}

We identify the key challenge of text-guided semantic editing as modifying visual content to align with the target prompt while preserving consistency with the original image. To address this challenge, in this work, we propose an efficient text-guided semantic editing method guided by latent space similarity (see in Fig.~\ref{fig1}). Since all operations are performed in the latent space, our method does not require access to high-dimensional internal model features and is compatible with both UNet-based and DiT-based architectures, making it a plug-and-play solution for text-guided semantic editing. In section \ref{adaptive_fusion}, we introduce the Adaptive Latent Fusion to achieve latent combination. Specifically, for a given source image $I^*$ and source prompt $P^*$, we first apply the image inversion technique to reverse $I^*$ to a specific noise sample $z_{T}^{*}$ and store the corresponding latent chain $\mathbf{z^*} = \{z_0^*, z_1^*, \cdots, z_t^*, \cdots, z_{T-1}^*, z_T^*\}$. This latent chain contains rich information about spatial layout, textural, and color features, which we incorporate into the inference process to effectively transfer the characteristics of the source image. Moreover, in section \ref{inversion_free}, we propose an inversion-free variant that approximates the intermediate reference latent $z_t^*$ following the forward process of diffusion models. This design makes it one of the most efficient methods that reduces NFEs by half while achieving performance comparable to SoTA methods.

\begin{figure}[t]
\includegraphics[width=\textwidth]{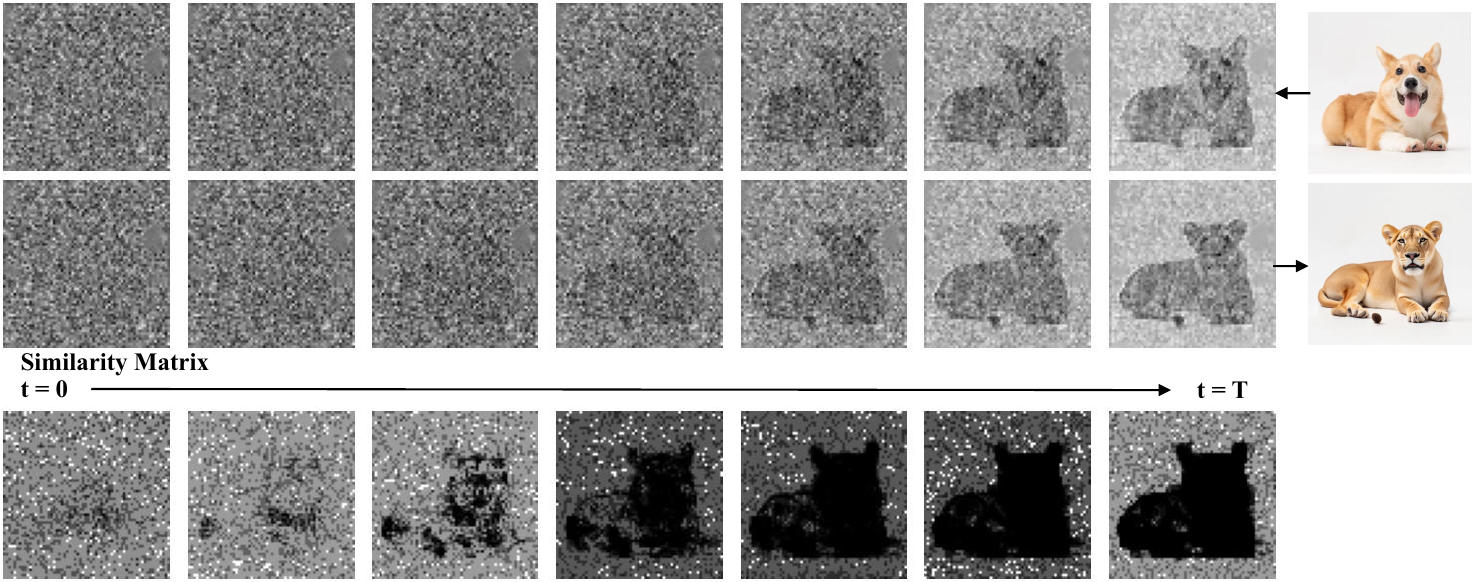}
\caption{Visualization of the reference latent chain, denoising latent states, and their similarity map.
Top to bottom: reference latent \(z_t^*\), current latent \(z_t\), and the similarity map between \(z_t\) and \(z_t^*\). The similarity guides selective feature preservation during editing.} 
\label{fig_vis}
\end{figure}

\subsection{Adaptive Latent Fusion}\label{adaptive_fusion}

\begin{algorithm}
\caption{Adaptivte Latent Fusion}
\begin{algorithmic}[1]
\State \textbf{Input:} Source prompt $P^*$ and source image $I^*$
\State \textbf{Output:} Target image $I$
\State Perform DDIM inversion or vanilla RF on $I^*$ to obtain latent trajectory $\{z^*_t\}_{t=0}^T$
\State Set initial latent $z_T \gets z^*_T$
\For{$t = T$ to $1$}
    \State Denoise $z_t$ to get $z_{t-1}$
    \State Compute mixed similarity $S_{\text{mix}} = \alpha \cdot \text{CosSim}(z_t, z^*_t) + (1 - \alpha) \cdot S_{\text{block}}$
    \State Compute final similarity map $S = \frac{1}{1 + \exp(-\gamma(S_{\text{mix}} - \tau))}$
    \State Fuse latents: $\hat{z}_t = z_t + S \odot (z^*_t - z_t)$
    \State Set $z_t \gets \hat{z}_t$
\EndFor
\State Decode $z_0$ to obtain edited image $I$
\State \Return $I$
\end{algorithmic}
\end{algorithm}

In previous approaches~\cite{Hertz2023P2P,Cao2023MasaCtrl,Tumanyan2023PnP,Wang2024RF-Solver,Deng2024FireFlow,Zhu2025KV-Edit}, researchers observed that the spatial layout, texture, and color of the generated images are influenced by the attention maps. Based on this observation, they attempted to inject the attention maps of the source image into the generation with the target prompt. However, directly injecting attention maps from the inversion process into the generation process may cause conflicts within the model, potentially leading to performance degradation. Furthermore, existing methods lack fine-grained control mechanisms, often resulting in unintended global or local background alterations in the generated images. In addition, they typically require storing a large number of high-dimensional attention features, which incurs substantial computational and memory overhead.

As illustrated in Fig.~\ref{fig_vis}, we observe that the latent space contains rich information that is highly correlated with the texture, edges, and spatial layout of the final generated image. This observation motivates us to guide the evolution of the latent space during the sampling process, allowing for effective injection of structural information from the source image. However, compared to attention maps, the latent representations themselves have a more direct impact on the final output. As a result, naively replacing the latent in the later steps of the denoising process would lead the generated image to overly rely on the source image, thereby undermining alignment with the target prompt and limiting generative flexibility.

To leverage the rich spatial information in the latent space while preserving the editability, we propose the adaptive latent fusion strategy that selectively incorporates spatial features from the source image at each timestep. 
% In particular, we fuse information primarily in background regions, while preserving only a minimal amount of original features in the foreground or semantically critical areas to maintain overall semantic consistency. 
Specifically, we first apply the inversion to the source image to obtain the source latent chain $\{ z_t^* \}_{t=0}^{T}$. We use $z_T^*$ as the initial noise sample for the denoising process. At each denoising timestep $t$, we compute the spatial similarity between the current latent $z_t$ and the corresponding inverted latent $z_t^*$. To capture pixel-level differences and model regional structural patterns, we propose a weighted similarity function that combines both channel-wise and block-wise similarities, which is defined as:
\begin{equation}
\mathbf{S}_{\text{mix}} = \alpha \cdot \text{CosSim}(\mathbf{z}_t^*, \mathbf{z}_t) + (1 - \alpha) \cdot \mathbf{S}_{\text{block}},
\label{eq:sim_mix}
\end{equation}
where $\alpha \in [0, 1]$ is a weighting factor that balances the trade-off between pixel-level precision and regional consistency. The block-wise similarity map $\mathbf{S}_{\text{block}}$ is calculated by dividing the spatial domain into non-overlapping blocks $B_{i,j}$, and computing the average cosine similarity within each block:
\begin{equation}
\mathbf{S}_{\text{block}}(i,j) = \frac{1}{|B_{i,j}|} \sum_{(u,v) \in B_{i,j}} \text{CosSim}\left( \mathbf{z}_t^*(u,v), \mathbf{z}_t(u,v) \right).
\label{eq:sim_block}
\end{equation}
Here, $\mathbf{S}_{\text{block}}(i, j)$ denotes the similarity score assigned to the spatial block located at the $(i, j)$ position. Each block is a non-overlapping region with a fixed size (\textit{e.g.}, $4 \times 4$) in the spatial dimensions of the latent feature maps. The terms $\mathbf{z}_t^*(u,v)$ and $\mathbf{z}_t(u,v)$ refer to the feature vectors at pixel position $(u,v)$ in their respective latent maps. The similarity is computed for each pixel pair within the block, and the average value over all pixels in $B_{i,j}$ is used to define the block-level similarity. This coarse-to-fine representation captures local semantic alignment and enhances robustness against noise and spatial distortions.

Since the raw similarity scores tend to be narrowly distributed, making it challenging to distinguish meaningful differences, we apply a non-linear transformation to enhance contrast and improve discriminability:
\begin{equation}
\mathbf{S} = \frac{1}{1 + \exp\left( -\gamma \cdot \left( \mathbf{S}_{\text{mix}} - \tau \right)\right)},
\label{eq: sigmoid}
\end{equation}
where $\gamma$ is a scaling factor, and $\tau = \mu + \lambda \cdot \left( \max(\mathbf{S}_{\text{mix}}) -\min(\mathbf{S}_{\text{mix}}) \right)$ is an adaptive threshold. Here, $\mu$ denotes the mean of $\mathbf{S}_{\text{mix}}$, and $\lambda$ controls the contribution of the dynamic range. This mapping enhances the similar regions while suppressing the distinct regions, thereby guiding more consistent feature blending in semantically important areas.Empirically, the search range of $\gamma$ is set to 20--200 and $\lambda$ to 0.04--0.12. With fewer inversion steps, where the discrepancy between inversion and generative noise is larger, a smaller $\gamma$ and larger $\lambda$ are preferred; with more inversion steps, $\gamma$ can be increased and $\lambda$ reduced to achieve a balanced trade-off.

Finally, we perform a weighted fusion of the current latent representation using the similarity map, enabling the selective incorporation of information from the source image:
\begin{equation}
\mathbf{{\hat{z}_t}} = \mathbf{z_t} + \mathbf{S} \odot (\mathbf{z_t^*} - \mathbf{z_t}),
\end{equation}
where $\odot$ denotes the Hadamard product. This formulation ensures that regions with high similarity retain more information from the source image, while regions with low similarity preserve the current latent, allowing better alignment with the target prompt. As a result, this blending mechanism preserves semantic consistency while enabling localized, controllable edits.

\subsection{Inversion-Free Semantic Image Editing}\label{inversion_free}
We further observed that the proposed method exhibits significant robustness to the quality of the inversion trajectory. Specifically, even in the absence of an accurate inversion of the source image, the editing results remain semantically coherent as long as the latent representations preserve sufficient spatial information from the source image. We posit that this robustness is due to the fact that, although the initial latent representation \( z_T \) may not be entirely accurate, it still contains adequate structural correspondence information. Through multiple iterations of refinement and optimization, the final latent representation accumulates sufficient structural similarity, which allows for the integration of a sufficient amount of necessary information from \( z_0 \) in the final generation step. This observation has motivated us to develop an inversion-free image editing approach to enhance efficiency.

A key challenge in inversion-free methods is retaining the spatial information of the source image, particularly in choosing an appropriate initial noise sample. Our observation indicates that directly using purely random noise as the initial seed often results in latent trajectories that lack spatial alignment with the source image, thereby limiting the effectiveness of our proposed fusion-guided mechanism. To address this issue, we construct the initial sample by the linear interpolation between the source image latent representation $z_0$ and a Gaussian noise sample $\epsilon \sim \mathcal{N}(0, \mathbf{I})$:
\begin{equation}
\mathbf{z}_T = \alpha \cdot \mathbf{z}_{\text{0}} + (1 - \alpha) \cdot \boldsymbol{\epsilon}.
\end{equation}
For the intermediate reference latent, we add noise to the image latent $z_0$ following the forward process of diffusion models. Specifically, we adopt different formulations for different models: for Stable Diffusion, we apply the DDIM-based deterministic forward process as defined in Eq.~(\ref{eq: DDIM add noise}); for FLUX, we follow the stochastic forward process with Rectified Flow as shown in Eq.~(\ref{eq: RF add noise}).

\section{Experimental Results}

In this section, we first conduct a comprehensive comparison with SoTA methods. Then, we present ablation studies to further demonstrate the effectiveness of the proposed method. See the appendix for more details and results.

\begin{table}[t]
\setlength{\tabcolsep}{3pt} 
\centering
\small
\caption{Comparison with SoTA methods on the PIE-Bench dataset.}
\label{tab:editing_comparison}
\begin{tabularx}{\textwidth}{Xccccccc}
\hline
\multirow{2}{*}{\textbf{Method}} & \multirow{2}{*}{\textbf{\begin{tabular}[c]{@{}c@{}}Structure\\ Distance$\downarrow$\end{tabular}}} & \multicolumn{2}{c}{\textbf{Fidelity}} & \multicolumn{2}{c}{\textbf{Editability}} & \multirow{2}{*}{\textbf{Steps}} & \multirow{2}{*}{\textbf{NFEs}} \\
 & & \textbf{PSNR}$\uparrow$ & \textbf{SSIM}$\uparrow$ & \textbf{Whole}$\uparrow$ & \textbf{Edited}$\uparrow$ & & \\
\hline
P2P~\cite{Hertz2023P2P} & 0.0699 & 17.84 & 0.7141 & 25.18 & 22.35 & 50 & 100 \\
MasaCtrl~\cite{Cao2023MasaCtrl} & 0.0276 & 22.36 & 0.8031 & 23.74 & 21.08 & 50 & 100 \\
PnP~\cite{Tumanyan2023PnP} & 0.0273 & 22.29 & 0.7934 & 25.21 & 22.46 & 50 & 100 \\
\cellcolor{Gray}{Ours} & \cellcolor{Gray}{0.0244} & \cellcolor{Gray}{23.09} & \cellcolor{Gray}{0.8016} & \cellcolor{Gray}{\textbf{25.67}} & \cellcolor{Gray}{\textbf{22.74}} & \cellcolor{Gray}{50} & 
\cellcolor{Gray}{100} \\
\cellcolor{Gray}{Ours} & \textbf{\cellcolor{Gray}{0.0224}} & \cellcolor{Gray}{\textbf{23.19}} & \cellcolor{Gray}{\textbf{0.8082}} & \cellcolor{Gray}{25.45} & \cellcolor{Gray}{22.51} & \cellcolor{Gray}{15} &  \cellcolor{Gray}{30} \\
\cellcolor{Gray}{Ours (Inv.-free)} & \cellcolor{Gray}0.0302 & \cellcolor{Gray}{22.73} & \cellcolor{Gray}{0.7972} & \cellcolor{Gray}{24.61} & \cellcolor{Gray}{21.68} & \cellcolor{Gray}{15} & \cellcolor{Gray}{15} \\
\hline
\multicolumn{8}{l}{\textit{RF-based methods}} \\
RF Inversion~\cite{Rout2025RF-Inversion} & 0.0446 & 20.31 & 0.7014 & 25.07 & \textbf{22.36} & 28 & 56 \\
RF-Solver~\cite{Wang2024RF-Solver} & 0.0332 & 22.69 & 0.8041 & 24.86 & 22.13 & 30 & 60 \\
FireFlow~\cite{Deng2024FireFlow} & 0.0288 & 22.87 & 0.8190 & 24.58 & 21.73 
& 8 & 18 \\
\cellcolor{Gray}{Ours} & \cellcolor{Gray}{0.0269} & \cellcolor{Gray}{23.12} & \cellcolor{Gray}{0.8178} & \cellcolor{Gray}{\textbf{25.28}} & \cellcolor{Gray}{22.14} & \cellcolor{Gray}{15} & \cellcolor{Gray}{30} \\
\cellcolor{Gray}{Ours} & \cellcolor{Gray}{\textbf{0.0265}} & \cellcolor{Gray}{\textbf{23.69}} & \cellcolor{Gray}{\textbf{0.8306}} & \cellcolor{Gray}{25.15} & \cellcolor{Gray}{21.90} & \cellcolor{Gray}{8} & \cellcolor{Gray}{16} \\
\cellcolor{Gray}{Ours (Inv.-free)} & \cellcolor{Gray}{0.2916} & \cellcolor{Gray}{22.86} & \cellcolor{Gray}{0.7845} & \cellcolor{Gray}{24.48} & \cellcolor{Gray}{21.71} & \cellcolor{Gray}{8} & \cellcolor{Gray}{8} \\
\hline
\end{tabularx}
\end{table}

\subsection{Comparisons with SoTA Methods}

\subsubsection{Quantitative Comparison.}
We conduct a comprehensive evaluation of the proposed method across different models on the PIE-Bench dataset~\cite{Ju2024PIEBench}. Quantitative results shown in Tab.~\ref{tab:editing_comparison} support the following two conclusions: (1) Regardless of DDIM-based or RF-based architecture, our proposed method consistently outperforms existing baselines in terms of background preservation and text-image alignment, while requiring significantly fewer denoising steps. (2) Our inversion-free variant achieves comparable performance to the SoTA methods, reducing NFEs by 50\% with only a 5-8\% drop of overall performance, making it well-suited for real-time applications.

\begin{figure}[t]
\includegraphics[width=\textwidth]{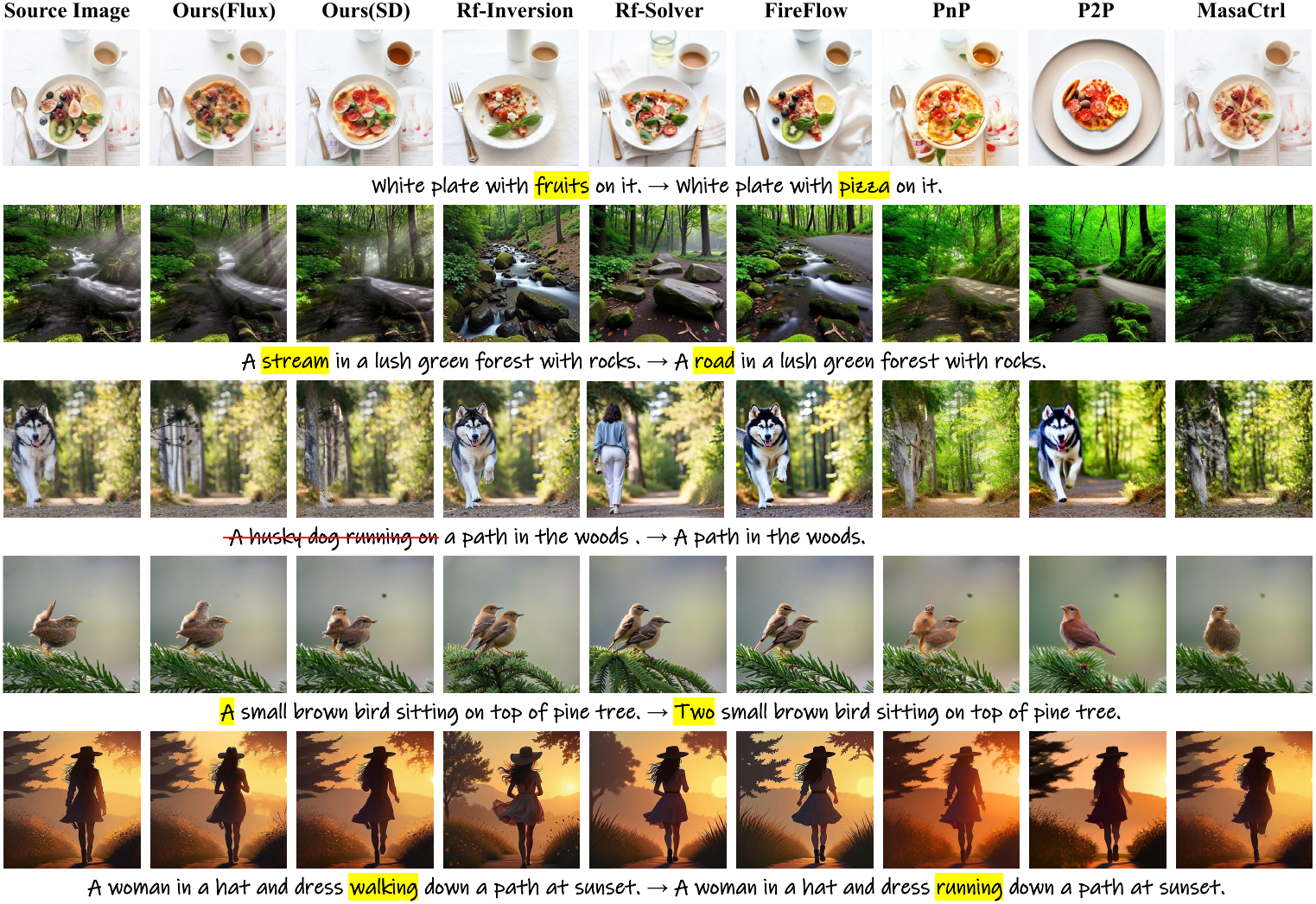}
\caption{Qualitative comparison with SoTA editing methods.
} 
\label{fig_re}
\end{figure}

\subsubsection{Qualitative Comparison.}
As shown in Fig.~\ref{fig_re}, our method exhibits a clear advantage in subjective comparisons. While DDIM-based approaches such as P2P\cite{Hertz2023P2P}, MasaCtrl\cite{Cao2023MasaCtrl}, and PnP\cite{Tumanyan2023PnP} are capable of effective editing, they often introduce excessive changes to unintended regions. RF-based methods like RF-Inversion\cite{Rout2025RF-Inversion}, RF-Solver\cite{Wang2024RF-Solver}, and FireFlow\cite{Deng2024FireFlow} mitigate this issue to some extent but still suffer from background inconsistencies. Moreover, both categories are prone to editing failures, as illustrated in the third and fourth rows of Fig.~\ref{fig_re}. In contrast, our approach achieves a better balance between fidelity and editability. It not only generates content that aligns accurately with the input text, but also maintains fine background details, resulting in superior overall visual quality.

\begin{figure}[t]
    \centering
    \label{fig:ab1}{%
        \includegraphics[width=\textwidth]{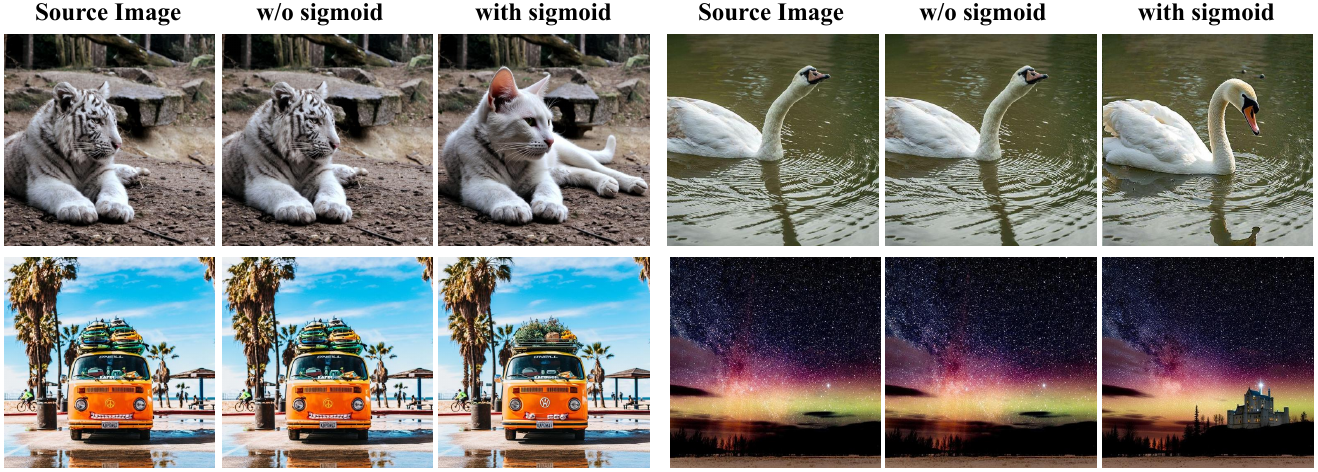}%
    }
    \text{(a) Ablation study on non-linear similarity transformation.}
    \\[0.2cm]
    \label{fig:ab2}{%
        \includegraphics[width=\textwidth]{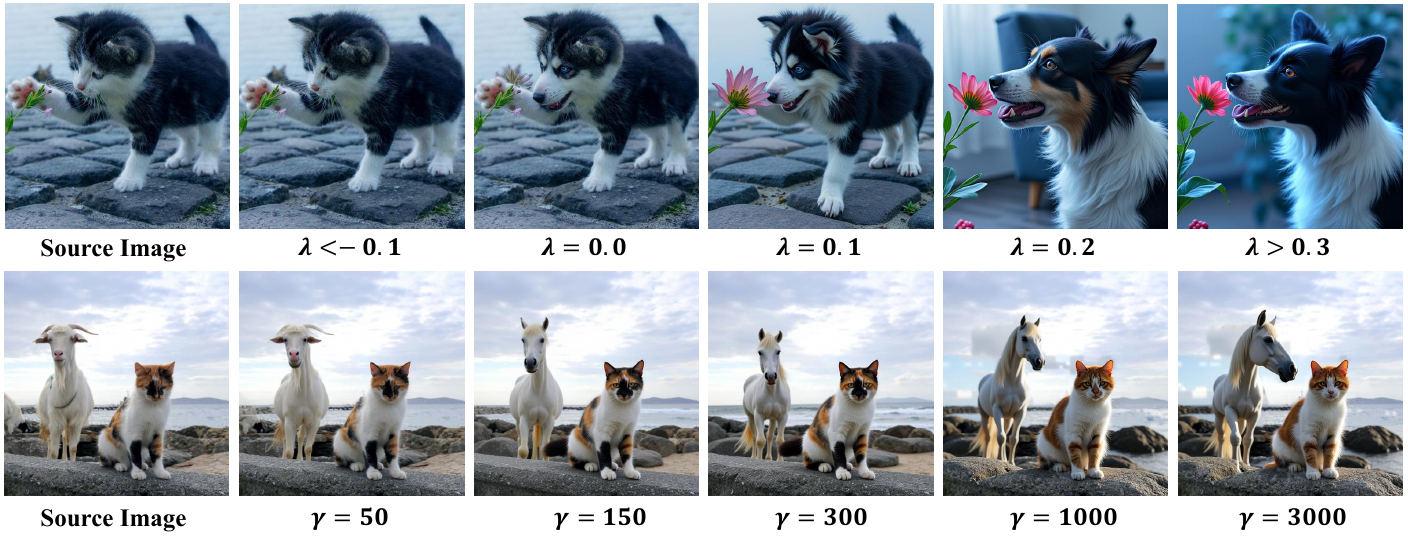}%
    }
    \text{(b) Ablation study on transformation parameters $\gamma$ and $\lambda$.}
    \caption{
Ablation studies on the similarity-guided fusion mechanism. 
(a) Effect of non-linear transformation: removing the sigmoid mapping leads to excessive preservation of the source image.
(b) Effect of transformation parameters. Both $\gamma$ (scaling factor) and $\lambda$ (dynamic range weight) control the strength and selectivity of feature blending. 
}
    \label{fig:ablation}
\end{figure}

\subsection{Ablation Studies}

\subsubsection{S vs $\mathbf{S}_{\text{mix}}$.}
We conducted an ablation study to evaluate the impact of the proposed nonlinear transformation applied to the similarity map. Specifically, we compared two feature fusion strategies: one using the original similarity map $\mathbf{S}_{\text{mix}}$ directly, and the other employing the similarity map $\mathbf{S}$ after applying a Sigmoid mapping as described in Eq.~(\ref{eq: sigmoid}). As shown in Fig.~\ref{fig:ablation}(a), without the nonlinear transformation, the similarity values exhibit minimal variation, making it difficult to distinguish regions that require editing from those that should be preserved. Consequently, the model tends to overly retain features, resulting in outputs nearly identical to the original images. In contrast, applying the Sigmoid transformation significantly enhances the contrast of the similarity map, clarifying semantic boundaries and yielding edited results that better align with the target prompt’s semantics, while effectively balancing semantic consistency and image fidelity.

\subsubsection{Hyperparameter in non-linear transformation formulation.} Fig.~\ref{fig:ablation}(b) further illustrates the impact of different hyperparameter settings in Eq.~\ref{eq: sigmoid}. Increasing the value of $\gamma$ significantly enhances the nonlinearity of the Sigmoid function, amplifying the numerical contrast between high- and low-similarity regions. This leads to stronger feature-blending effects. Raising $\lambda$ increases the adaptive threshold $\tau$, which results in more regions being classified as low similarity, thereby suppressing background feature retention. Experiments show that both $\gamma$ and $\lambda$ contribute to greater editing flexibility at the expense of semantic consistency. Notably, $\gamma$ has a more significant influence on the final image outcome and exhibits higher sensitivity compared to $\lambda$.

\begin{table}[t]
\setlength{\tabcolsep}{3pt} 
\centering
\small
\caption{Ablation study on the block size.}
\label{tab:ablation}
\begin{tabular}{cccccc}
\hline
\multirow{2}{*}{\textbf{Block Size}} & \multirow{2}{*}{\textbf{\begin{tabular}[c]{@{}c@{}}Structure\\ Distance$\downarrow$\end{tabular}}} & \multicolumn{2}{c}{\textbf{Fidelity}} & \multicolumn{2}{c}{\textbf{Editability}} \\
 & & \textbf{PSNR}$\uparrow$ & \textbf{SSIM}$\uparrow$ & \textbf{Whole}$\uparrow$ & \textbf{Edited}$\uparrow$ \\
\hline
1  & 0.0368 & 21.75 & 0.7965 & 25.55 & 22.42 \\
2  & 0.0304 & 22.92 & 0.8168 & 25.32 & 22.14 \\
4  & 0.0265 & 23.69 & 0.8306 & 25.15 & 21.90 \\
8  & 0.0237 & 24.34 & 0.8411 & 25.01 & 21.82 \\
16 & 0.0211 & 24.95 & 0.8481 & 24.81 & 21.65 \\
32 & 0.0192 & 25.48 & 0.8515 & 24.64 & 21.54 \\
\hline
\end{tabular}
\end{table}

\subsubsection{Block Size.} We further investigate the impact of the block size in the proposed block-wise similarity module. As shown in Table~\ref{tab:ablation}, larger block sizes consistently yield lower structure distance as well as higher PSNR and SSIM, indicating improved structural consistency and pixel-level fidelity. However, the CLIP score decreases as block size increases, suggesting a trade-off between visual fidelity and semantic alignment. We attribute this phenomenon to the receptive field of block-wise similarity: when the block size is larger, the similarity measure emphasizes broader regional consistency and suppresses local noise, thus enhancing pixel-level reconstruction quality. Conversely, finer block sizes focus more on localized alignment, which benefits semantic correspondence captured by CLIP, but may lead to noisier pixel-level reconstructions. This observation highlights the importance of balancing block size in order to achieve the desired trade-off between fidelity and semantic faithfulness.

\section{Conclusion}
We propose \textbf{LatentEdit}, a novel and efficient framework for consistent semantic image editing that operates directly in the latent space. By leveraging adaptive latent fusion guided by spatial similarity between the denoising latent and a reference latent chain, LatentEdit enables precise control over feature preservation and prompt-driven modifications. Different from prior methods that rely on high-dimensional internal features, our approach avoids model conflicts and memory overhead, offering a plug-and-play solution compatible with both UNet-based and DiT-based architectures. Our proposed method also includes an inversion-free variant that approximates reference latents via forward diffusion, reducing NFEs by half. Extensive experiments on the PIE-Bench dataset demonstrate that the proposed method achieves SoTA performance in both fidelity and editability, with significantly lower computational cost.

\section*{Acknowledgments}

This work is supported by the National Natural Science Foundation of China (No. 62331014) and Project 2021JC02X103. We acknowledge the computational support of the Center for Computational Science and Engineering at Southern University of Science and Technology.

% \bibliography{ref}

% ---- Appendix ----

\section*{Appendix}

\section*{A. Experimental Settings}

\subsubsection{Baselines.}
We conduct the experiment across two baselines: SD v1.5 with DDIM sampler and FLUX.1-dev with RF sampler (Euler sampler). Besides, we compare our method with DDIM training-free image editing approaches: P2P\cite{Hertz2023P2P}, MasaCtrl\cite{Cao2023MasaCtrl}, and PnP\cite{Tumanyan2023PnP}. We also consider the recent RF inversion methods, such as RF-Inversion\cite{Rout2025RF-Inversion}, RF-Solver\cite{Wang2024RF-Solver}, and FireFlow\cite{Rout2025RF-Inversion}.

\subsubsection{Implementation Details.}
Since our method is plug-and-play, we conduct experiments on both SD v1.5 and FLUX.1-dev to validate its effectiveness. For the FLUX model, we perform image editing tasks using 8 and 15 steps, with guidance scales set to 1.5 for the inversion process and 3.5 for the denoising process. For the Stable Diffusion model, we first invert the image into the initial noise map using deterministic DDIM inversion~\cite{Song2021DDIM,Dhariwal2021DDIM}. The classifier-free guidance~\cite{Ho2021CFG} scale is set to 1.0 during inversion. During the denoising process, we apply DDIM sampling with 50 and 15 denoising steps, using a guidance scale of 5.5. Other baselines retain their default parameters or use previously published results. All experiments are conducted on a single NVIDIA L40 GPU, and the resolution of all test images was set to 512 × 512.

\subsubsection{Evaluation Metrics.}
To ensure a fair comparison, we evaluate our method and baselines on the PIE-Bench. The dataset consists of 700 images with 10 types of editing, where each image is paired with a source prompt and a target prompt. To evaluate our method and other baselines, we use seven metrics across three dimensions: text-guided quality, preservation quality, and time cost. For background preservation, we measure Structure Distance~\cite{Ju2024PIEBench}, Peak Signal-to-Noise Ratio (PSNR), and Structural Similarity Index (SSIM)~\cite{Wang2004SSIM}. For text-image alignment, we report the CLIP~\cite{Radford2021CLIP} score.

\begin{algorithm}
\caption{Adaptive Latent Fusion (Inversion-Free)}
\begin{algorithmic}[1]
\State \textbf{Input:} Source prompt $P^*$, target prompt $P$, and source image $I^*$
\State \textbf{Output:} Target image $I$
\State Encode $I^*$ to obtain latent $z_0$
\State Sample noise $\epsilon \sim \mathcal{N}(0, I)$
\State Set $z_T = \alpha \cdot z_0 + (1 - \alpha) \cdot \epsilon$
\State Generate pseudo-reference latents $\{z^*_t\}_{t=0}^T$ via forward process
\For{$t = T$ to $1$}
    \State Denoise $z_t$ with target prompt $P$ to get $z_{t-1}$
    \State Compute mixed similarity $S_{\text{mix}} = \alpha \cdot \text{CosSim}(z_t, z^*_t) + (1 - \alpha) \cdot S_{\text{block}}$
    \State Compute final similarity map $S = \frac{1}{1 + \exp(-\gamma(S_{\text{mix}} - \tau))}$
    \State Fuse latents: $\hat{z}_t = z_t + S \odot (z^*_t - z_t)$
    \State Set $z_t \gets \hat{z}_t$
\EndFor
\State Decode $z_0$ to obtain edited image $I$
\State \Return $I$
\end{algorithmic}
\end{algorithm}

\section*{B. More Results of Image Editing}

\subsubsection*{Inversion-Free Image Editing.}
To validate the effectiveness of our approach, we present inversion-free editing results in the supplementary materials. As shown in Fig.~\ref{fig:if}, our method is capable of generating content that aligns closely with the text description while faithfully preserving the original background, all without relying on latent space inversion. This clearly demonstrates the core strengths of our method in achieving both semantic precision and structural integrity, highlighting that its effectiveness is not dependent on inversion techniques.

To further aid understanding of our approach, we also provide pseudocode for the inversion-free editing process in this section, offering a clear illustration of its implementation details.

\begin{figure}
    \centering
    \label{fig:if-sd}{%
        \includegraphics[width=\textwidth]{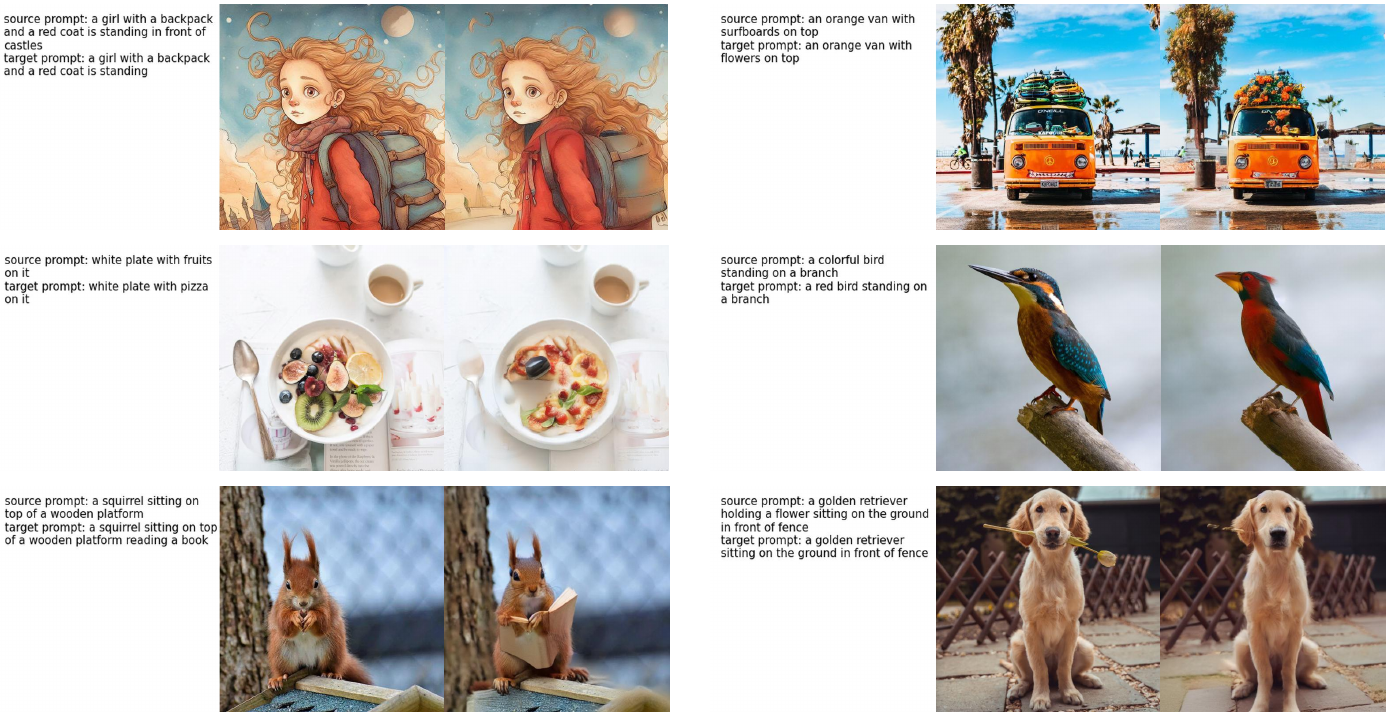}%
    }
    \text{(a) Inversion-free semantic image editing results with Stable Diffusion.}
    \\[0.2cm]
    \label{fig:if-flux}{%
        \includegraphics[width=\textwidth]{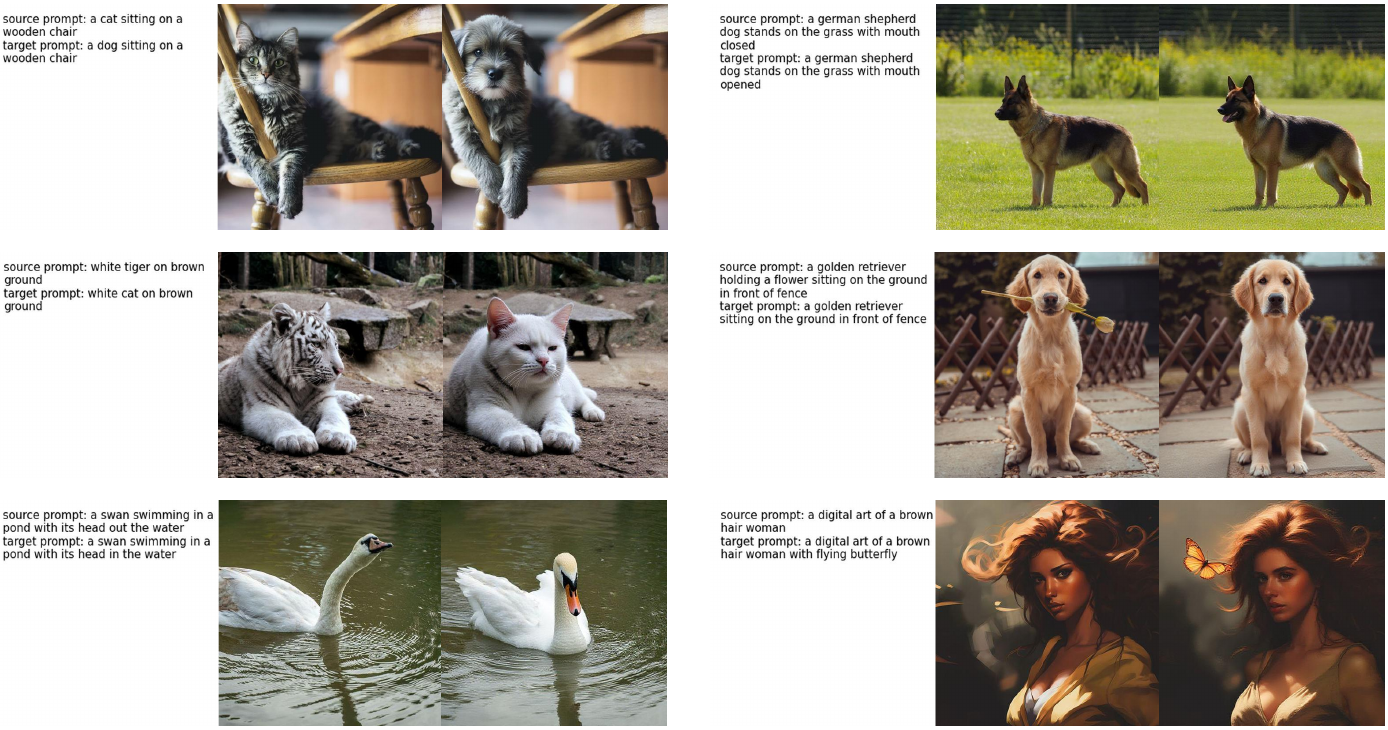}%
    }
    \text{(b) Inversion-free semantic image editing results with FLUX.}
    \caption{Results of inversion-free semantic image editing. Our method achieves effective prompt-driven edits while preserving background details, without requiring explicit latent inversion. Each group of images is organized as follows: the first column shows the source prompt and target prompt; the second column displays the source image generated from the source prompt; the third column presents the edited target image corresponding to the target prompt. 
    }
    \label{fig:if}
\end{figure}

\subsubsection*{Failure Cases.}

\begin{figure}
\includegraphics[width=\textwidth]{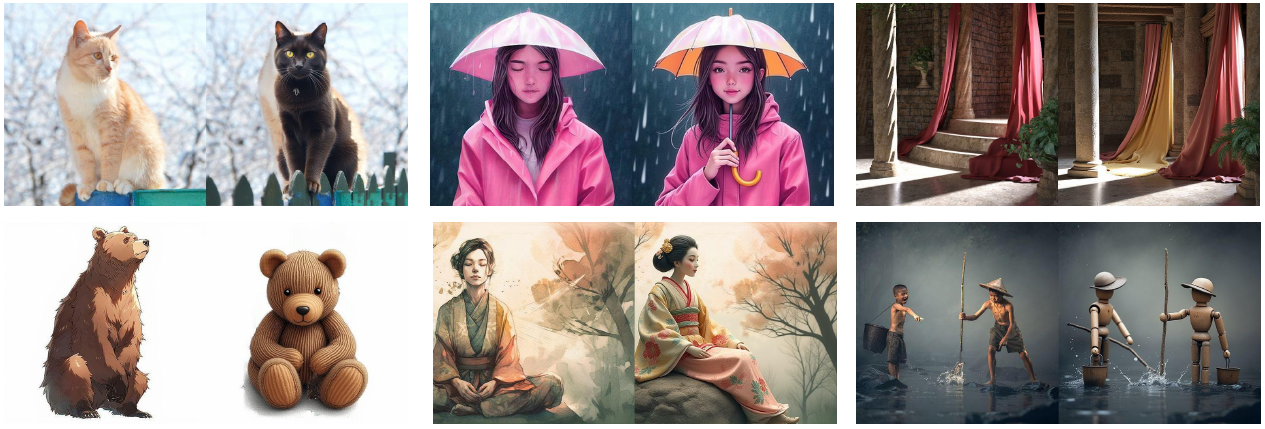}
\caption{Illustrations of failure cases in semantic image editing. These examples highlight typical scenarios where our method struggles, particularly when editing subtle attributes such as color or material. In each group, the first column shows the source image, while the second column displays the target image resulting from a failed editing attempt. The results demonstrate how unintended changes to key visual features can occur alongside the desired edits.}
\label{fig:failures}
\end{figure}

We empirically observe that our method often fails when attempting to edit subtle attributes of the main subject in an image, such as its color or material. In these cases, modifying one attribute tends to unintentionally alter or degrade other key features of the subject. As shown in Fig.~\ref{fig:failures}, the first row demonstrates a failed attempt to change the object's color, resulting in noticeable distortion of its original appearance. The second row illustrates a failure in editing the material, where the intended modification compromises the structural integrity or identity of the subject.

\section*{C. Disscussion}
While our method demonstrates strong performance on a variety of editing tasks, we observe notable limitations when it comes to modifying subtle attributes of the main subject in an image, such as color or material. Empirical results indicate that such fine-grained edits often lead to unintended alterations in other key visual features, sometimes even compromising the identity of the subject. For instance, attempts to change the object's color or material may simultaneously distort shape, texture, or other defining characteristics.

We hypothesize that these failures stem from the limited granularity of control imposed by operating at a relatively low-resolution latent space. At such a scale, the model lacks the capacity to isolate and manipulate fine attributes without affecting the broader semantic representation of the image. In other words, the entanglement of attributes in the latent space leads to over-coupled changes during editing.

To address this issue, future work could explore performing adaptive fusion directly within the attention layers of the model. This would potentially allow for more precise and disentangled control over various image attributes, enabling more targeted modifications without compromising global coherence or subject identity.

\end{document}